*Chapter 1*

# HOW DO EYES AND BRAIN SEARCH A RANDOMLY STRUCTURED UNINFORMATIVE SCENE? EXPLOITING A BASIC INTERPLAY OF ATTENTION AND MEMORY

*L. Resca*[*1], *P. M. Greenwood*[2] *and T. D. Keech*[1]

[1]Physics Department, The Catholic University of America, Washington, DC
[2]Psychology Department, George Mason University, Fairfax, VA

**ABSTRACT**

[*] Corresponding author: Prof. Lorenzo Resca, Physics Department, The Catholic University of America, Washington, D.C. 20064. Tel: (202) 319-5334. Fax: (202) 319-4448. E: resca@cua.edu

We tracked the eye movements of seven young and seven older adults performing a conjunctive visual search task similar to that performed by two highly trained monkeys in an original influential study of Motter and Belky (1998a, 1998b). We obtained results consistent with theirs regarding elements of perception, selection, attention and object recognition, but we found a much greater role played by long-range memory. A design inadequacy in the original Motter-Belky study is not sufficient to explain such discrepancy, nor is the high level of training of their monkeys. Perhaps monkeys and humans do *not* use mnemonic resources compatibly already in basic visual search tasks, contrary to a common expectation, further supported by cortical representation studies. We also found age-related differences in various measures of eye movements, consistently indicating slightly reduced conspicuity areas for the older adults, hence, correspondingly reduced processing and memory capacities. However, because of sample size and age differential limitations, statistically significant differences were found only for a few variables, most notably overall reaction times. Results reported here provide the basis for demonstrating the formation of spiraling or circulating patterns in the eye movement trajectories and for developing corresponding computational models and simulations.

## 1. INTRODUCTION

The ability to visually search a typically complex and confounding environment for designated or relevant targets is critical to most animal life, and it has correspondingly evolved into extremely complex and powerful detection systems. Starting with basic sensors, eyes in higher animals, including humans, have developed a progressively increasing acuity from the periphery to the central region of the retina or fovea. A basic mechanism of visual search has correspondingly developed. That consists of rapid deployments of the eyes fovea to potential target locations, punctuated by relatively longer eye fixations, during which items at, or near the fovea are processed for target recognition. During each fixation, subsequent eye movements are also programmed, so that the search may proceed if the target is not recognized.

A number of questions naturally arise from such basic observations. What guides or attracts the eyes from one fixation to another? How close or spread consecutive eye fixations may be and why? To what extent visual search patterns may be random, preset, influenced by the scene appearance or characteristic of the viewing individual? Which features and items are examined during each eye fixation and how and to what extent are those processed before the eyes may move elsewhere? Which or how much information may be retained, transferred or integrated across eye movements, for how long, and how may that be used to structure the visual search and make it more efficient?

A great deal has been learned about those and other major questions regarding visual search through many psychophysics and neuroscience studies, vastly increasing in number and sophistication in recent years. Seminal studies revealed how attention may guide visual search (Treisman and Gelade, 1980; Wolfe, 1994) and eye movements (Zelinsky and Sheinberg, 1997). Other landmark studies indicated how memory may or may not be concurrently involved (Horowitz and Wolfe, 1998; Klein, 1988; Klein and MacInnes, 1999; Wolfe, 2003). Search strategies and scanpaths in eye movements were also found and originally characterized (Noton and Stark, 1971; Stark and Choi, 1996). Among many other subsequent studies, particularly notable are those of Peterson, Kramer, Wang, Irwin, and McCarley (2001), Peterson, Beck, and Vomela (2007), and Dickinson and Zelinsky (2007), who further characterized retrospective and prospective memory in visual search, or a high-capacity but low-resolution memory for the search path.

A particularly influential study on active visual search was originally conducted by Motter and Belky (1998a, 1998b), who demonstrated the critical role played by a zone of focal attention in guiding eye movements. A basic intent of that study was to display visual scenes that have barely enough structure to induce serial search with eye movements ―thus called *active* visual search― but otherwise offer only minimal guidance. So, Motter and Belky trained extensively rhesus monkeys to perform a paradigmatic visual search task in which the target is a bar item characterized by a unique *conjunction* of *color* and *orientation*, placed somewhat randomly among distractor bars that share with the target, in equal proportions, either one of those "visual primitive features" (Treisman and Gelade, 1980). Naturally, the eye-brain system is tuned to searching scenes with much more complex structure, thus eliciting both greater guidance and confounding. So, the purpose of that prototypical task is to reveal constituents of the eye-brain system that operate at a most basic level. Those constituents provide basic tools that the system can further adapt and develop to guide or inform visual search when additional structure and confounds are progressively imposed on the scene (Henderson, Chanceaux, and Smith, 2009).

Based on extensive investigations of eye movements and fixations, Motter and Belky summarized and integrated their conclusions with the following model. (1) Recognition of the target typically occurs with high probability only within a restricted area surrounding the current fixation location. The size of this zone of attentional focus, called a*rea of conspicuity* (AC) according to Engel (1971, 1974, 1977), decreases with increasing *density* of *relevant stimuli*. (2) If the target is not detected within the current AC, a saccade is made most likely to an item just beyond that AC. That item has most likely the same *color* as the target, rather than the same *orientation*. (3) Except for those constraints, the item providing the new fixation point is selected apparently at *random*. (4) A *forward bias* is nevertheless apparent in consecutive saccades.

## 1.1. Correcting a Design Inadequacy of the Motter-Belky Study

The results and model of Motter and Belky provided critical information on how the AC operates precisely as to guide and constrain eye movements at a fundamental level. On the other hand, we noticed an undesirable feature in the experimental settings of Motter and Belky that could have had some unintended consequences. It derived from a design artifact with the placement of the target in their displays. Namely, Motter and Belky (1998a) generated a fixed set of 44 target locations on their screen as follows: "Target locations were equidistantly spaced along imaginary concentric rings (six per ring) occurring at increments of 2 deg from screen center. The target location in each successively larger ring were rotated 20 deg clockwise, giving an overall spiral configuration" (p. 1008). That structure may organize the displays to some degree even after random selection of target and distractor locations. Highly trained monkeys could have learned implicitly and taken advantage of that underlying structure procedurally after having performed so many trials. Even more importantly, the *radial* probability density of placing the target in an annulus with radius $r$ from the screen center and infinitesimal thickness $dr$ is constant in such design, rather than proportional to $r$. Correspondingly, the target location *density* is inversely proportional to $r$, rather than constant, which is instead required of a truly random and *uniform* distribution of target locations throughout the display. Thus, monkeys were presented with search arrays where the target was more likely placed in central than peripheral regions, in inverse proportion to the distance $r$ from the screen center. Since the monkeys fixated the screen center at the beginning of each trial, they were thus provided with an artificially greater chance of capturing the target promptly. Conversely, when they did not detect the target quickly and kept moving their eyes outward, the monkeys were handicapped with an artificially smaller chance of capturing the target in those peripheral regions. This could have affected a proper assessment of any sort of long-range memory, a feature that is prominently excluded in the Motter-Belky model.

Given the fundamental importance of the Motter-Belky study for any further understanding and modeling of eye movements in visual search, we designed a similar study aimed initially at eliminating their design inadequacy, thus confirming and extending their main results and model more validly and reliably. Subsequently, our investigation has led to a study of pattern formation in eye movement trajectories by Keech and Resca (2010a), who demonstrated spiraling characteristics of those movements by considering autocorrelations and power spectra of saccade directions and by introducing certain measures of circulation and subtended area in the eye trajectories projected on the display. Our investigation has also led to the development of a computational model and corresponding computer simulations that require the presence of long-range memory to confirm the formation of spiraling patterns in the observed eye trajectories and to account for their return rates (Keech and Resca, 2010b).

In this chapter, we first recompile the experimental results of our original investigation, expand their observations and describe their conclusions. We thus correct the design inadequacy of the Motter-Belky study, and yet we confirm or complement many of their original results and observations. This is critical, because of the major impact that the original papers of Motter and Belky (1998a, 1998b) have had in the field and the influence that they continue to exert.

## 1.2. Human Participants vs. Trained Monkeys

Secondly, we address the issue that, while our visual search task and psychophysical measurements were quite similar to those of the main experiment of Motter and Belky (1998a, 1998b), there was a major difference regarding participants. Namely, while Motter and Belky used two highly trained rhesus monkeys as their subjects, we worked with fourteen human participants much less practiced with the visual search task. Differences that we found in corresponding observations may have further implications and far reaching consequences that we shall discuss.

### 1.3. Long-Range Memory

Thirdly, and perhaps most importantly, this chapter concentrates on our investigation and conclusions about a long-range memory component, based on analyses and observations of "survivor functions" and rates of return in the eye trajectories to previously inspected locations. That also demonstrates, beyond the Motter-Belky model, the intertwined roles that the AC plays in both the *identification* of objects and their *memory tagging* to avoid re-inspection ―which may be capsized by saying that one cannot "remember" as a non-target what has not been first "identified" as a non-target (whether correctly or incorrectly).

Several important studies have begun to reveal the immense complexity and adaptability of the eye-brain system in its ability to analyze and interpret internally a complex and continuously changing visual environment, depending on the task at hand (Ballard, Hayhoe, Pook, and Rao, 1997; Boccignone and Ferraro, 2004; Brockmann and Geisel, 2000; Najemnik and Geisler, 2005; Navalpakkam and Itti, 2007; Peters, Iyer, Itti, and Koch, 2005; Privitera and Stark, 2000; Stephen, Mirman, Magnuson, and Dixon, 2009; Torralba, Oliva, Castelhano, and Henderson, 2006). From that perspective, one could imagine that elements or traces of complex search strategies or biases would persist even in visual searches of *uninformative scenes* with *randomized stimuli*. That is not apparently the case, according to the Motter-Belky model and some findings of ours. Namely, for the displays and task of our experiments, Keech and Resca (2010b) have been able to model phenomenologically and simulate computationally the search movements of the eyes almost as if they belonged to a "mechanical" system, subject only to relatively simple rules and procedures such as those of the Motter-Belky model. However, that does not mean that the acting of eye-brain system is not particularly "smart" or "intelligent" at times. To the contrary, it is at least "smart" enough to realize when to deploy or not to deploy which more or less sophisticated and costly resources it has in its arsenal to optimize the efficiency of the visual search. For our visual search task, we know by design that the only critical element that can make the search more efficient is to use some basic combination of an attentional and a long-range memory component. The *long-range memory* component, far exceeding current estimates of a rapidly decaying *inhibition of return* (IOR; Snyder and Kingstone, 2000), is where the Motter-Belky model needs critical improvement, as Keech and Resca (2010a, 2010b) have demonstrated and we shall further illustrate in this chapter. Other authors (Aks, Zelinsky, and Sprott, 2002) have also analyzed power spectra of eye fixation series in a task (of finding the upright "T") relatively similar to ours and concluded that long-range memory across eye movements in the form of complex self-organizing search patterns is used to provide maximum coverage of a search area with minimal effort.

Most recent studies of eye movements while viewing naturalistic scenes have revealed further intricacies in the structure and organization of IOR and its cortical representations. On the one hand, Smith and Henderson (2009, 2011a, 2011b) have shown that latencies of saccades landing close to the immediately previous (one-back) fixation location become elevated, consistently with oculomotor IOR. However, the likelihood that such previous location is re-fixated is not diminished, contrary to the IOR supposition. Still, an overall forward bias of saccades is maintained (Klein and MacInnes, 1999; MacInnes and Klein, 2003) or reframed as a "saccadic momentum" (Smith and Henderson, 2009, 2011a, 2011b) or an "attentional momentum" (Spalek and Hammad, 2004). By devising more complex probabilistic analyses and simulations of gaze patterns recorded during both free-viewing and search of naturalistic scenes, Bays and Usain (2012) may have resolved at least some of those apparent contradictions by showing that gaze history has a much greater influence than other accounts suggesting that visual search of natural scenes has no memory.

### 1.4. Age effects

There is a fourth aspect in the design of our study, addressing the question of how aging may affect various elements and mechanisms of the visual search process. That is, of course, a very important question, which has been considered in many previous studies ―see, for instance, Ball, Beard, Roenker, Miller, and Griggs (1988); Castel, Chasteen, Scialfa, and Pratt (2003); Davis, Fujava, and Shikano (2002); Davis, Shikano, Peterson, and Michel (2003); McCarley, Kramer, Colcombe, and Scialfa (2004); Scialfa (2002); Scialfa, Kline, and Lyman (1987); Sekuler and Ball (1986). Age defined comparisons of eye movements have also been studied in particular ―see, for instance, Dennis, Scialfa, and Ho (2004); Ho and Scialfa (2003); Scialfa, Jenkins, Hamaluk, and Skaloud (2000); Scialfa, Thomas, and Joffe (1994).

We planned to look into this question by comparing a group of 7 young adults, aged 18 to 28, with an average age of 20.4 years and a standard deviation of 3.5 years, with a group of 7 older adults, aged 42 to 55, with an average age of 49.7 years and a standard deviation of 5.8 years. Although there is a substantial statistical difference between the age means of those two groups ($t_{[12]} = -11.4$, with $p < 10^{-7}$), our older adults may not be considered sufficiently "old" by typical standards of many studies on aging. Correspondingly, we did not typically find either major or definitively significant differences in visual search characteristics between our two age groups. Nevertheless, subtle changes or decline in many cognitive functions often begin at relatively young ages, while remaining unremarkable or unnoticed for many more subsequent years. So, it is quite worthwhile in view of a longer perspective on research and diagnostic developments to investigate and look for subtle changes or early signs of cognitive decline or aging, such as those underlying many of our observations on visual search.

Thus, after collecting data from each individual performing the same task under the same conditions, we performed statistical analyses separately for each age group with regard to all the eye movement variables, the types of trials, the correlation, circulation and area measures in the eye trajectories, and the corresponding computer simulations that we devised. We found, for example, that the older adults need slightly more saccades to detect the target, and correspondingly incur slightly larger failure rates. Those and many other consistently related results convinced us that older adults have slightly reduced conspicuity areas and, correspondingly, slightly less processing and memory capacities. Although still small, such deficits are quite consistent with the age-related "generalized slowing hypothesis", which likely affects visual search tasks (Plude and Doussard-Roosevelt, 1989; Scialfa and Joffe, 1997) like many other cognitive skills and abilities. Indeed, we found ourselves significantly longer *reaction times* to detect targets in the older age group for displays of all but the largest array size, having $t_{[12]} = -2.161, -3.932, -2.120$, and $-1.250$, with $p = 0.052, 0.002, 0.056$, and $0.235$, for 12-, 24-, 48-, and 96-distractor arrays, respectively.

On the other hand, we found that the age differential between our two groups and their sample size were too limited in order to technically attain statistical significance for most other differences in separate visual search characteristics. While recognizing that our study is thus underpowered in that regard, we have opted nonetheless to analyze and report most of our results and figures separately for each age group, so that we may still notice and point out some suggestive or insightful differences and comparisons. Failure to reach statistical significance according to conventional standards does not imply that any particular observation is necessarily meaningless or uninformative. Such failure may even be valuable and interesting in and of itself ―perhaps "comforting" growing older individuals! So, we believe and maintain that at least some trends that we have observed are real, consistent and informative. At a minimum, our study indicates what sample sizes may be needed in larger future studies of eye movements to provide firmer evidence of age-related differences in certain visual search characteristics that could be critical for both diagnostic and practical implementations.

## 2. METHODS

Participants in our investigation were 7 young adults, aged 18-28, with an average age of 20.4 years, and 7 older adults, aged 42-55, with an average age of 49.7 years.[1] Participants wore a head-piece holding an Applied Science Laboratories (ASL) Model 501 eye-tracker and a magnetic head tracker. The eye-tracker sampled at rate of 240Hz, corresponding to a sampling interval of 4.1667 ms. Participants were seated at an average distance of 26 inches from the center of a 17-in. Macintosh video display with effective dimensions of $11.25 \times 7.5$ in. This results in the search array being evenly distributed across a field of view of about $24.4° \times 16.3°$. The accuracy of our fixation measurements corresponds to a visual angle of about $0.5°$. Motter and Belky (1998a, 1998b) had chronically implanted scleral search coils on their two rhesus monkeys, allowing a considerably greater visual-angle accuracy of about $0.05°$ to $0.1°$. The field of view in the Motter-Belky main experiment (standard display) was $34° \times 25.5°$, thus carrying a density of objects lower than in our experiment by a 2.2 factor, approximately.

The search arrays in our investigation were developed generating *randomly* (with a pseudorandom number generator) the X and Y Cartesian coordinates of the centers of all the objects (distractors and target) in the display, with the only constraint of a minimum center-to-center separation of $1.4°$. This allows relative density fluctuations to decay with the inverse square root of the number of objects in the display, which is typical of random "molecular" collections. On the other hand, Motter and Belky (1998a, 1998b) covered their display area with a square lattice and distributed distractors randomly in the lattice cells, which produces slightly more uniform distributions. However, the placement of the *target* in the Motter-Belky main experiment was neither completely random nor uniform, but rather orderly and preferentially *central*, which could have undermined certain basic assumptions of their statistical analysis of target capture probabilities. This design problem will be discussed in detail in Sec. 4.2. Our participants searched for a target bar that was randomly placed among 12, 24, 48, 96 additional distractor bars, in 90 *trials* for each array size. Interspersed randomly among those, there were 9 additional trials for each array size in which the target was absent.

---

[1] All our experiments were undertaken with the understanding and written consent of each participant and all our procedures and protocols were approved by the Institutional Review Board. Participants were free to discontinue their participation at any time and confidentiality was granted. The authors had no conflicts of interest.

Participants were not informed of the number of target-absent trials.[2] Before starting the actual experiment, each participant practiced briefly for 20 trials, comprising 5 trials for each array size.

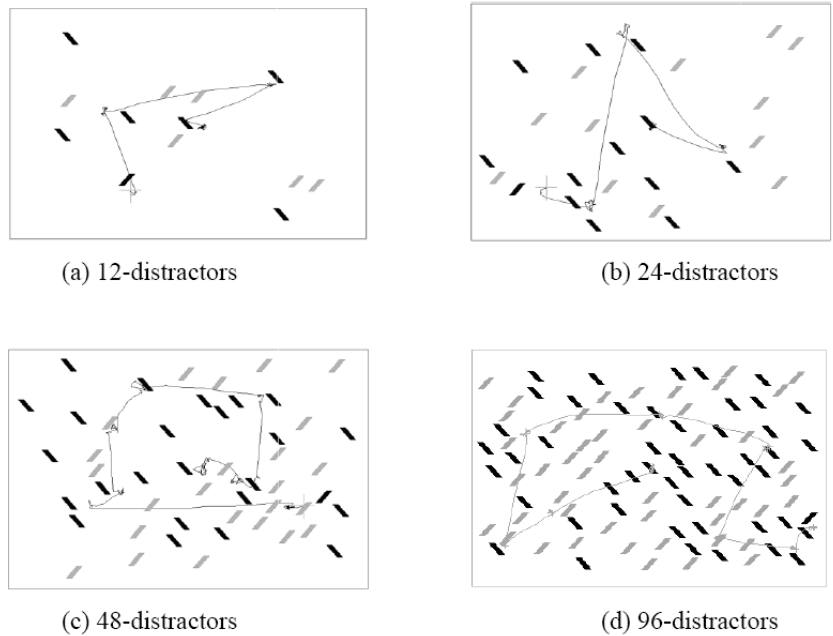

(a) 12-distractors          (b) 24-distractors

(c) 48-distractors          (d) 96-distractors

Figure 1. Examples of search arrays with: (a) 12, (b) 24, (c) 48, (d) 96 distractors. Actual eye-trajectories of an observer are superimposed. The target is characterized by a unique conjunction of color and orientation: black bar, tilted $45^{o}$ to the right. Half of the distractors share color with the target but have a different orientation (tilted $45^{o}$ to

---

[2] The 9:1 target present/absent ratio was reasonably chosen to discourage participants from prematurely guessing or falsely reporting the presence of targets, and still collect a greater number of target-present trials, since only those could be used later for comparison with computer simulations (Keech, 2006; Keech and Resca, 2010b).

the left), while half of the distractors shared orientation with the target but differ in color (red, appearing as grey in these figures). A cross indicates the point at which the target was detected by the observer.

The target was characterized by a unique conjunction of color and orientation: black bar, tilted $45^o$ to the right. Half of the distractors shared color with the target but had a different orientation (tilted $45^o$ to the left), while half of the distractors shared orientation with the target but differed in color (red). Since the display stood approximately 26 inches away from the eyes of the participant, all bars subtend an angular length of 0.7° and an angular width of 0.15°, approximately. The bar size in the Motter-Belky main experiment was $1.0^o \times 0.25^o$, which is larger than ours by a 2.4 factor, thus scaling approximately as the display size.

Participants were instructed to fixate a cross appearing at the *center* of the display at the beginning of each trial. Beyond that initial fixation, no further instruction was given with regard to eye movements, in order to allow for any spontaneous use of the eyes in response to the task demands. On the other hand, participants were required to make a speeded search decision about the presence or absence of the target in the array by *pressing* one of two buttons on a keypad with their left or right hand, respectively. The search array remained on for 4.0 sec or until a response to the search task was made. By contrast, Motter and Belky (1998a, 1998b) trained their monkeys to *fixate* for at least 600 ms the target, always present, in order to complete the trial. Their search array stayed on for 7.262 sec or until the target was detected.

Figure 1 shows four examples of our search arrays, with actual eye movement trajectories superimposed.

## 3. MEASUREMENTS

In this section we present the measured frequency distributions of the number of saccades per trial, the durations of the fixations and their distances from the nearest distractors or the detected target, and the saccade amplitudes, durations, velocities and changes of direction. Reaction times and target recognition accuracy are also reported. A discussion of our treatment of measurement errors is provided in an Appendix.

### 3.1. Saccade Number Distributions

For each array size, Figures 2(a-d) show the frequency distributions of the number of saccades in target-present trials, averaged over 7 young and 7 older adults, separately. The mean and the standard deviation of each distribution are reported in the figure captions. Corresponding distributions for young and older adults are relatively similar, although young adults need slightly fewer saccades to detect the target than the older adults do, on average. However, differences in the average number of saccades per trial are not statistically significant between those two relatively small groups.

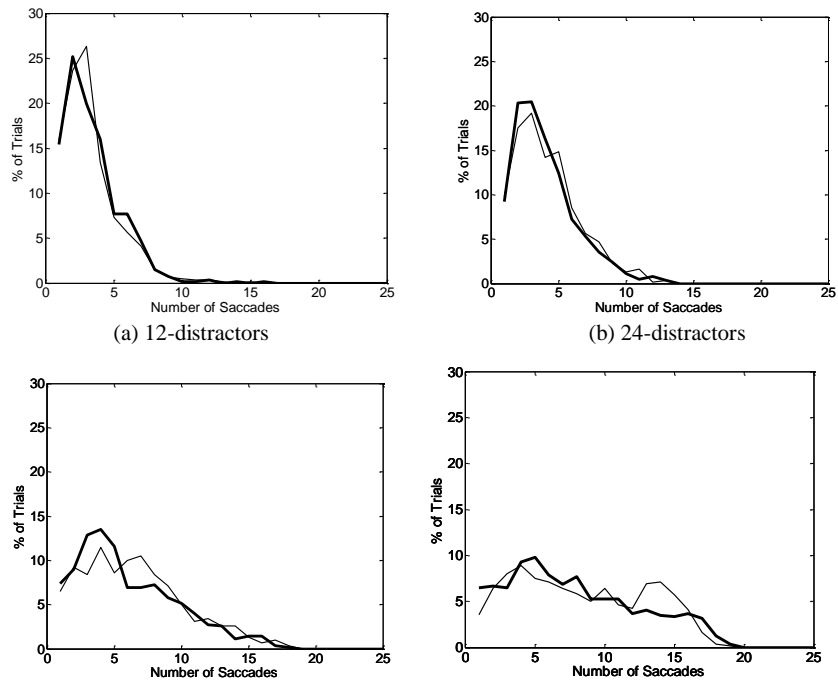

Figure 2. Frequency distributions (percentages of trials) of the number of saccades per trial for young (thick lines) and older (thin lines) adults. For each search array size, the Mean/Standard Deviation for young (old) are: (a) 3.4/2.1 (3.3/2.0); (b) 4.0/2.3 (4.2/2.4); (c) 6.1/3.7 (6.5/3.8); (d) 7.8/4.7 (8.4/4.6).

Figure 3 plots the average number of saccades per trial (with target present) vs. search array size for young and older adults. A linear increase in the average number of saccades with the search array size is typically expected and associated with serial processing in conjunctive visual search (Treisman and Gelade, 1980; Zelinsky and Sheinberg, 1997). A basically linear increase is indeed displayed in both our age groups, apart from a slight reduction for the 96-distractor arrays. That reduction is partly a consequence of the 4-second time limit imposed on the search, which reduces the successful searches to those with relatively fewer saccades. Namely, while participants pressed the left button indicating target detection in virtually all search arrays with a target and 12 and 24 distractors, a few participants failed to press any button for arrays with 48 distractors, and considerably more participants failed likewise for arrays with 96 distractors. Specifically, the failure rates for the 48- and the 96-distractor arrays were 0.96% and 7.75% for the young adults, and 2.12% and 15.12% for the older adults. The failure rate difference between young and older participants is not statistically significant for 48-distractor arrays ($t_{[12]} = -1.284$, $p = 0.62$) and borderline significant for 96-distractor arrays ($t_{[12]} = -2.114$, $p = 0.056$).

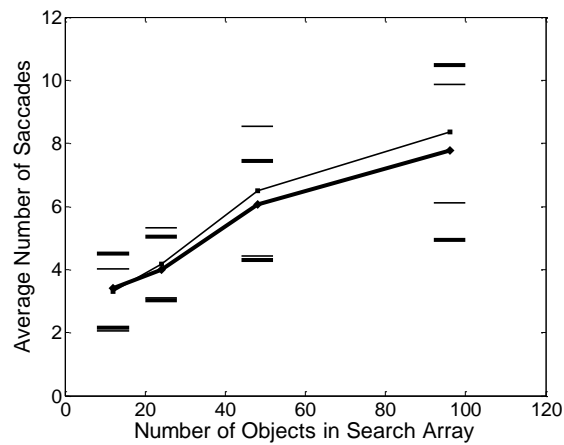

Figure 3. Averages (over 90 search trials) for young (thick line) and older (thin line) participants of the number of saccades per trial, as a function of search array size. Extreme individual values are indicated with thick (thin) bars for young (older) participants.

On the other hand, Figure 15 of Motter and Belky (1998a, p. 1018) shows a quite similar reduction in the linear increase of the average number of saccades, even though their monkeys had practically no time limit, hence, virtually no failure rate for detecting the target. Notice that the *average numbers* of saccades of our human participants and the monkeys performing the "standard" task in Figure 15 of Motter and Belky (1998a, p. 1018) appear to be similar. However, we will show at the end of Sec. 4.1 that the complete *frequency distributions* of the number of saccades actually differ substantially in humans and monkeys, and we will explain why.

The saccade number distributions that we have just presented play a central role in the construction of a computational model of the search movements of the eyes and serve as a critical benchmark for comparison with the corresponding computer simulations (Keech and Resca, 2010b).

### 3.2. Fixation Duration Distributions

Figure 4 shows the fixation duration distributions for young and older adults, with their means and standard deviations reported in the captions, merging data for all the array sizes. The initial fixation duration is typically longer and is not included in the distributions shown in Figure 4, nor in Table 1, which lists the average durations of mid-trial fixations for young and older adults searching each array size separately.

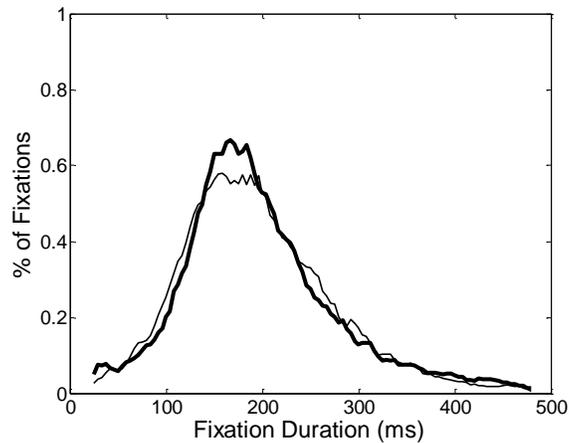

Figure 4. Fixation duration distributions (percentages of fixations per ms) for young (thick line) and older (thin line) adults. Mean/Standard Deviation for young (old) adults are: 196.2/80.0 (194.9/77.8).

Those averages are nearly identical for young and older adults, and increase only slightly with array size. Specifically, the overall increase in the average fixation duration between the value for the 12-distractor array and the value for the 96-distractor array is approximately 12% for young adults and 16% for the older adults. Consequently, fixation duration is nearly independent of the local density of objects. These results are thus consistent with those of Motter and Belky (1998a, Figure 4, p. 1011) and their conclusion that items are processed in parallel within the AC. So, unlike the number of saccades, fixation durations are not diagnostic of serial searching, but rather reflect stimulus factors in more complex ways (Zelinsky and Sheinberg, 1997). We have also determined that the duration of a fixation is statistically independent of the angle-shift between its preceding and following saccade.

**Table 1. Average durations (ms) of midtrial fixations for 12, 24, 48, and 96-distractor arrays**

| Participants | Search Array Size | | | |
|---|---|---|---|---|
| | 12 | 24 | 48 | 96 |
| Young Adults | 184 ms | 185 ms | 196 ms | 206 ms |
| Older Adults | 178 ms | 183 ms | 196 ms | 206 ms |

**Table 2. Average durations (ms) of the initial fixation for 12, 24, 48, and 96-distractor arrays**

| Participants | **Search Array Size** | | | |
|---|---|---|---|---|
| | 12 | 24 | 48 | 96 |
| Young Adults | 344 ms | 326 ms | 325 ms | 308 ms |
| Older Adults | 368 ms | 359 ms | 361ms | 365 ms |

Table 2 shows the average durations of the initial fixation for young and older adults searching each array size. Those durations are relatively shorter for young than older adults, with the difference reaching a level of statistical significance for the largest array size. Namely, we have $t_{[12]} = -1.11, -1.64, -1.57, -2.49$, with $p = 0.29, 0.13, 0.14, 0.029$, for 12-, 24-, 48-, 96-distractor arrays, respectively. Motter and Belky (1998a) have suggested some processes that may differ during the initial and subsequent fixations in a subsection on "initial vs. mid-trial search performance" (p. 1020-1021). However, the initial fixation was typically no longer than subsequent ones in their highly trained monkeys, which is definitely not the case in our human participants.

### 3.3. Reaction Time

For each search array size, Table 3 shows the means and standard deviations of the average reaction time (RT) distributions for young and older adults. The standard deviations of the RT distributions are relatively large (about half the average RT values).

**Table 3. Means and Standard Deviations of the average reaction time (RT) distributions for young and older participants (ms)**

| Search Array Size | Young Adults RT | | Older Adults RT | |
|---|---|---|---|---|
| | Mean | Standard Deviation | Mean | Standard Deviation |
| 12 | 834 | 336 | 1036 | 439 |
| 24 | 1016 | 470 | 1290 | 547 |
| 48 | 1592 | 775 | 1878 | 859 |
| 96 | 2240 | 1040 | 2429 | 1121 |

However, in order to compare the two age groups, we have to consider for each array size the mean $<X>$ of the means $<X>_i$ of the RT distributions for each of the i=1,..,7 young adults, and the mean $<x>$ of the means $<x>_j$ of the RT distributions for each of the j=1,..,7 older adults. We then have to consider the (altogether different and much smaller) standard deviation statistics S and s of $<X>_i$ and $<x>_j$ from $<X>$ and $<x>$, respectively. We can use those statistics to form $t_{[12]} = (<X> - <x>)/(S^2/7 + s^2/7)^{1/2}$, which we can compare with the usual student's t-distribution with 12 degrees of freedom. From that we can conclude that, on average, young adults capture the target significantly faster than older adults, except for the 96-distractor arrays, since $t_{[12]}$ = -2.161, -3.932, -2.120, -1.250, with $p$ = 0.052, 0.002, 0.056, 0.235, for 12-, 24-, 48-, 96-distractor arrays.[3]

---

[3] We should mention that in the RT distributions just discussed we included also trials that ended in failure to press any button, assigning the maximum of 4 seconds as RT for the trials that actually timed out. We did that in order to account to the maximum extent possible for the fact that older adults had greater failure rates for the larger arrays (see Sec. 3.1). Excluding timed-out trials, RT distributions change somewhat for the larger arrays, yielding somewhat

Although the RT is an all-encompassing measure, it seems to be mostly affected differentially in young and older adults by the number of fixations and by the duration of the initial fixation. Although those differences did not appear to reach statistical significance separately, they probably combined to produce that in the RT aggregate. However, other components may have also contributed to that effect. Since we are mainly interested in specific measures of eye movements and fixations, rather than in the global measure of RT, which is correspondingly more "confounded" (Aks, 2009, p. 135), we have not tried to further parse the RT and determine more precisely the origins of its significant differential in our two age groups.[4]

## 3.4. Saccade Amplitude Distributions

Saccade amplitude distributions are central to our study and we need to begin their discussion with a basic classification and operational definition of related types of eye movements.

The brain can keep the eyes practically still by storing in memory their position through persistent neural activity integrated in the brainstem and cerebellum (Seung, 1996). Nevertheless, a certain level of *tremor* is unavoidable, involving random eye movements with high frequencies (30-70 Hz) and small amplitudes (less than $0.5°$). Correspondingly, eye tremor velocities typically do not exceed $35°/s$. *Slow control* refers to involuntarily drifting smooth motion of the eyes (with a typical velocity of about $0.1°/s$) in an attempt to fixate a stationary target. Saccade velocities far exceed the velocities of either the slow control of eye movements or the *smooth eye pursuit* of moving objects. Saccades are further distinguished from eye tremor by larger amplitudes (exceeding $0.5°$) and greater velocities (exceeding $35°/s$). So, while our eye-tracker was quite capable of detecting and discriminating all these types of eye movements, which did indeed appear in our recorded data and eye trajectories, we simplified their analysis and classification by using

---

different values than those reported in Table 3. However, young-old statistical differences remain similar when timed-out trials are excluded.

[4] We should also point out that we used multiple t-tests for variables that in principle may not be all orthogonal. In that case, however, any more elaborate factorial ANOVA would only confirm the lack of statistical significance or power that we have already found for most individual factors. On the other hand, our t-test for statistical significance of the difference between young and older adults in overall RT still indicates that that effect is real, since RT could have been measured exclusively and independently (as in many other studies on visual search) and a single t-test could have been performed correspondingly.

$40^{o}$/s as a velocity threshold below which a *fixation* was declared and above which a *saccade* was declared, according to a relatively common practice and standard in the literature (Sperling, 1990).

Figure 5 shows the saccade amplitude distribution for the older adults searching 48-distractor arrays. Notice that this distribution exhibits a sharp peak at short saccades, with amplitudes less than one degree. The same feature appears prominently in Figures 5 and 6 of Motter and Belky (1998b, p. 1809).

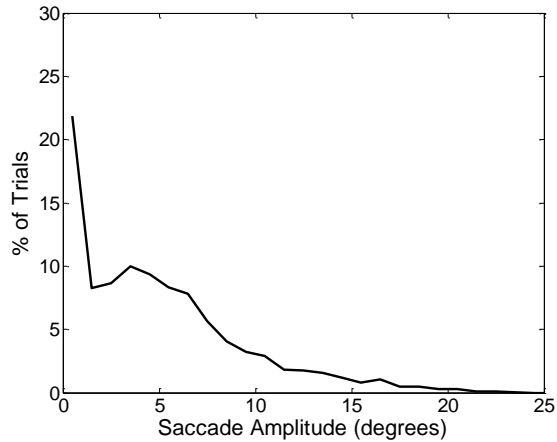

Figure 5. Saccade amplitude distribution (percentage of saccades per degree of amplitude) for the older adults searching 48-distractor arrays.

Visual search is mainly conducted during periods of relatively long eye fixations (~200 ms), followed by saccades typically lasting for about 30 ms, which bring the fovea to bear on a different region of space. In examining the relation between the saccade amplitude and the duration of the preceding fixation, we found that saccades with small amplitudes (less than $1^{o}$) tend to be precipitous, i.e., they are typically preceded by short fixations (less than 5 ms). Furthermore, consider Figure 6 for example, which shows a scatter plot of the shift in saccade angle, relative to the preceding saccade angle, for cases in which the intervening fixation lasts less than 10 ms and the ensuing saccade has a plotted amplitude of less than $5^{o}$. Notice that the greatest concentration of scatter points (44% for older adults and 47% in a similar plot for young adults) occurs between $0.4^{o}$ and $1.3^{o}$ degrees for the ensuing saccade amplitude and between $140^{o}$ and $220^{o}$ degrees for its angle change. This indicates that small-amplitude and precipitous saccades typically revert direction relative to their preceding saccades.

These observations suggest that small-amplitude saccades are inessential to the visual search process and mainly provide some immediate correction to a previous saccade overshoot. Motter and Belky (1998b, p. 1808-1809) have previously reached similar conclusions. We thus decided to exclude from further processing and considerations saccades with amplitudes smaller than 1.3$^o$. Although Motter and Belky did not make any such decision, one should keep in mind that there is actually no discrepancy between our results and theirs in that regard.

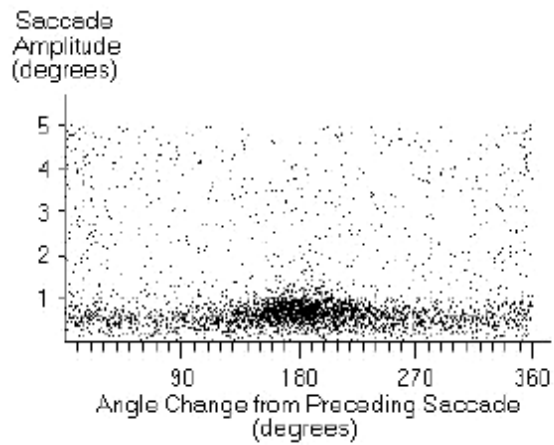

Figure 6. Scatter plot for saccades with amplitudes of less than 5$^o$ and any angle-shift relative to their preceding saccades, but with an intervening fixation duration of less than 10 ms. Data are for the older adults, merging all array sizes.

Saccade amplitude distributions, with the restriction just mentioned, are shown in Figures 7(a-d) for young and older adults searching each array size. These distributions are relatively similar for the two age groups, although the most probable amplitudes are slightly shorter for the older adults, suggesting slightly smaller conspicuity areas for the older adults. Furthermore, if we superimpose these distributions for all array sizes in either age group, there is still considerable overlap, as indicated by all their relatively close means. Vice versa, a decreasing trend in the means is noticeable for increasing array size in the corresponding Figure 5 of Motter and Belky (1998b, p. 1809). This apparent discrepancy may result from the slight difference in *uniformity* between displays in the two studies.

Ours exhibit a slightly greater clustering, resulting from greater density fluctuations, especially for the smaller array sizes ─observe, for instance, our Figures 1, compared to Figure 1 of Motter and Belky (1998a, p. 1008). Because of such clustering, there are slightly more small amplitude saccades for the smaller search array sizes in our Figures 7 than in Figure 5 of Motter and Belky (1998b, p. 1809). This trend discrepancy still appears, although less noticeably, if we compare our Figures 7(a-d) with Figure 9.6 of a subsequent article by Motter and Holsapple (2001, p. 167).

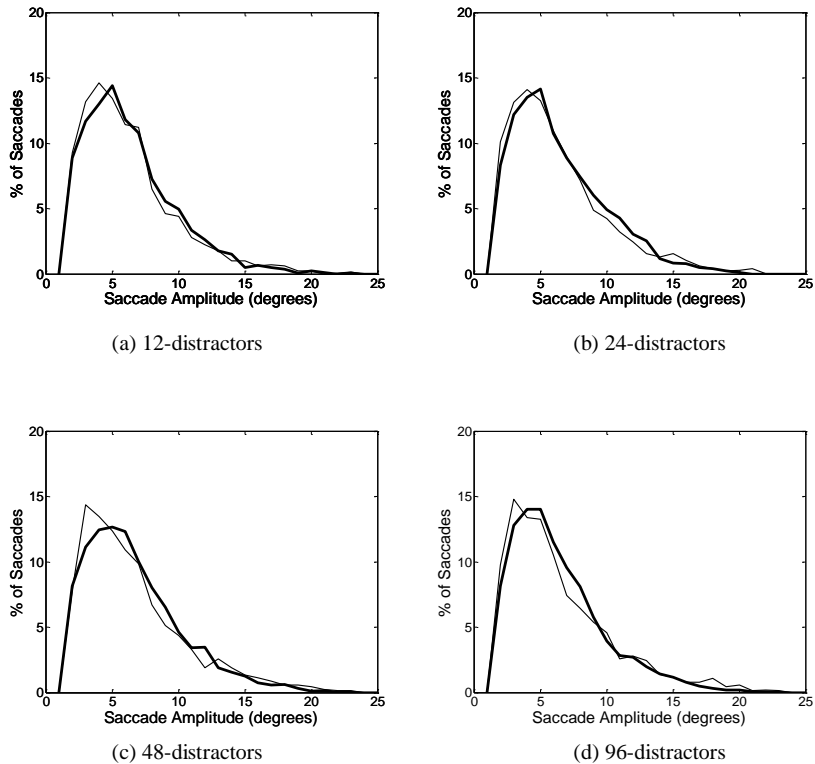

(a) 12-distractors

(b) 24-distractors

(c) 48-distractors

(d) 96-distractors

Figure 7. Saccade amplitude distributions (percentages of saccades per degree of amplitude) for young (thick lines) and older (thin lines) adults. For each search array size, the Mean/Standard Deviation for young (old) are: (a) 6.35/3.39 (6.24/3.52); (b) 6.46/3.45 (6.36/3.73); (c) 6.65/3.57 (6.64/3.92); (d) 6.36/3.45 (6.50/3.98).

Nevertheless, this trend discrepancy is not intrinsically meaningful, and it can actually be removed as follows. Consider the average distance between *relevant* nearest neighbors (NN), i.e., those having the same *color* (or c-) as the target. This average NN distance, or c-ANND, is $4.9°$, $3.6°$, $2.7°$, and $2.1°$ for the 12-, 24-, 48-, and 96-distractor arrays, respectively, and it is determined as follows. For each individual array, the distance from each black object to all other black objects is computed to determine the c-NND to each black object. These c-nearest neighbor distances are then averaged over all black objects in each search array. The procedure is then repeated for all search arrays of the same size and averaged over those, thus yielding the corresponding c-ANND.

Expressing amplitudes in c-ANND units, the saccade amplitude distributions are plotted again in Figures 8(a-d) for young and older adults. The peaks and means of these distributions now shift to larger c-ANND multiples with increasing array size for all sizes.

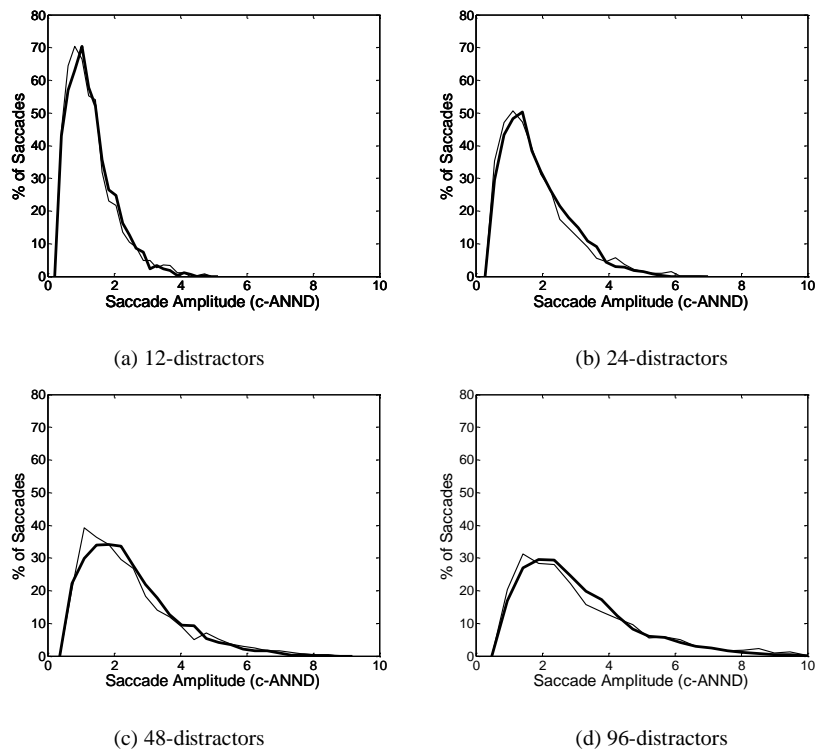

(a) 12-distractors
(b) 24-distractors
(c) 48-distractors
(d) 96-distractors

Figure 8. Saccade amplitude distributions for young (thick lines) and older (thin lines) adults in c-ANND units, which are: (a) $4.9^o$, (b) $3.6^o$, (c) $2.7^o$, and (d) $2.1^o$. For each search array size, the Mean/Standard Deviation for young (old) are: (a) 1.31/0.69 (1.29/0.72); (b) 1.82/0.96 (1.80/1.04); (c) 2.46/1.31 (2.46/1.44); (d) 3.02/1.62 (3.09/1.88).

Thus, using c-ANND units has the expected effect of compensating the greater clustering effect for the smaller array sizes. These results then become consistent with all those reported in Figure 6(B) of Motter and Belky (1998b, p. 1809). The saccade amplitude distributions that we have reported include all trials for each array size. However, as also noted by Motter and Belky (1998b), there are significant differences among the distributions of the first saccade, mid-trial saccades, and the last saccade. Measured saccade amplitude distributions for each of those types of saccades were fully reported elsewhere (Keech, 2006) and then used to generate saccade probability factors in computer simulations (Keech and Resca, 2010b, Figures 2 and 3).

### 3.5. Distributions of Distance from Fixation Location to Nearest Distractor

Figures 9 show the distributions of the distances from the current fixation to (a) the nearest *relevant* distractor, sharing the same color (black) with the target, and (b) the nearest *irrelevant* (red) distractor. Measured data for search arrays of all sizes are combined, since there is little difference between distributions for those separately —even though the ANND is much greater in smaller than in larger arrays, namely, $3.6^o$, $2.8^o$, $2.1^o$, $1.6^o$, for 12-, 24-, 48-, 94-distractor arrays. These distributions show that saccades land much closer to relevant than to irrelevant distractors —slightly more so in young than in older adults. Our distributions are thus essentially similar to those reported in Figures 2 of Motter and Belky (1998b, p. 1807) for their two monkeys —even though the monkey distributions involving the irrelevant distractors are much flatter than ours.

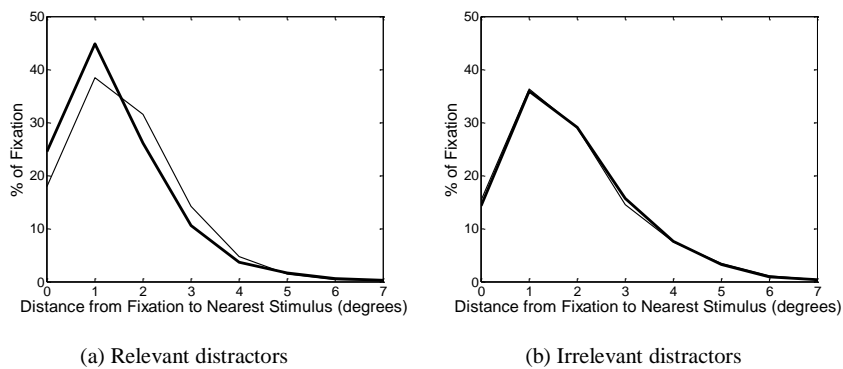

(a) Relevant distractors    (b) Irrelevant distractors

Figure 9. Distributions (percentages of fixations per degree) of the distances from the current fixation to the center of the nearest relevant (black) distractor (a) and irrelevant

(red) distractor (b) for young (thick lines) and older (thin lines) adults. The Mean/Standard Deviation for young (old) are: (a) 1.57/1.15 (1.74/1.12); (b) 1.96/1.27 (1.95/1.30).

These results confirm that saccades target preferentially stimuli of the same *color*, rather than the same *orientation* ―a key point (2) in the Motter-Belky model. The corresponding *color segmentation* (or *filtering*) has figured prominently since the earliest studies in *serial conjunctive search* (Egeth, Virzi, and Garbart, 1984; Pashler, 1987; Treisman and Gelade, 1980) and it is, of course, a central element in the computational model of Keech and Resca (2010b).

## 3.6. Relative Amplitudes and Change of Direction in Consecutive Saccades

Figure 10(a) shows a scatter plot of the amplitudes of consecutive saccades for young adults, while Figure 10(b) shows a scatter plot of the angle change in the direction of consecutive saccades. In both figures, final saccades are excluded and data from all search arrays are combined. Plots for the older adults are similar (Keech, 2006). In Figure 10(b), the radial distance represents the ratio of the amplitude of the following saccade to the amplitude of the preceding saccade. These plots are similar to those in Figure 7 of Motter and Belky (1998b, p. 1810) ―although there is some correlational tendency for shorter saccades to be followed by longer saccades in Figure 7(A) of Motter and Belky (1998b, p. 1810), which is not apparent in our Figure 10(a).

Remarkably, Figure 7(B) of Motter and Belky (1998b, p. 1810) shows a slight bias for the next saccade to avoid the area just crossed by the preceding saccade. A return saccade to the preceding fixation is also notable therein for about 3% to 4% of all saccades.[5] Thus, Motter and Belky (1998b) unquestionably found a bias to execute consecutive saccades in a consistent direction ―plus a small tendency to return to the item just previously fixated, perhaps for the purpose of verification. A *forward bias* in consecutive saccades is equally notable in our Figure 10(b). This forward bias is a basic point (4) of the Motter-Belky model and a central element for the development of spiraling patterns in the eye movement trajectories (Keech and Resca, 2010a, 2010b). As a psycho-neural mechanism, it may be related to IOR

---

[5] The degree labels in Figure 7(B) of Motter and Belky (1998b, p. 1810) appear to have been drawn upside down.

(Klein and MacInnes,1999; MacInnes and Klein, 2003) or other forms of saccadic momentum (Smith and Henderson, 2009, 2011a, 2011b) or attentional momentum (Spalek and Hammad, 2004).

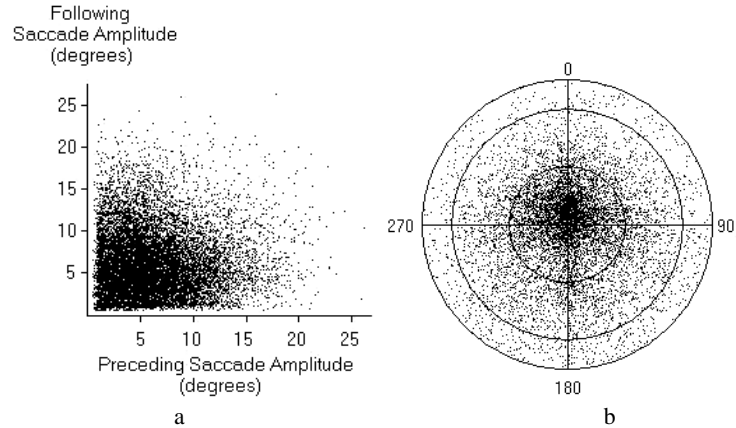

Figure 10. Scatter plots of consecutive saccade amplitudes (a) and angle change in degrees (b) for young adults, combining data from all search arrays. In (b), the radial distance from the origin represents the ratio of the amplitudes of the next and the previous saccade, with circles marking ratios of 1, 2 and 3.

Figure 11, obtained from young adults searching 48-distractor arrays, shows a scatter plot of the shift in saccade angle relative to the preceding saccade angle for the cases in which the intervening fixation duration is longer than 10 ms and both saccades have amplitudes between $1.4^o$ and $15^o$. Corresponding plots for both young and older adults searching arrays of all sizes are similar to this Figure 11 (Keech, 2006).

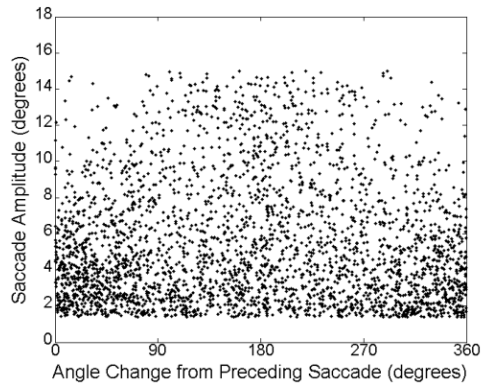

Figure 11. Scatter plot of amplitude and angle-shift for saccades with a preceding fixation duration longer than 10 ms. Data are for young adults searching 48-distractor arrays.

The concentration of points shows that the angle-shifts occur preferentially in the *forward* direction (less than $90^o$ or more than $270^o$) when the second saccade has an amplitude of less than about $7^o$, but almost equally in the *backward* direction (between $90^o$ and $270^o$) when the second saccade has an amplitude of more than about $7^o$. The former trend suggests again a *bias* for consecutive saccades to progress in a *forward* direction, while the latter trend is related to *reversals* in saccade directions occurring more frequently at the display *boundary* for larger saccade amplitudes.

These observations provide critical support for the interpretation of saccade direction autocorrelations in particular (Keech and Resca, 2010a, Figure 1, p. 120). They also provide an empirical basis for the formulation of an angle-shift dependence factor in the saccade probabilities of the computational model and simulations of Keech and Resca (2010b).

### 3.7. Saccade Velocity and Duration Distributions

Figure 12 shows the saccade velocity distribution, averaged over all search array sizes, for young and older adults. Most saccades have velocities between $90^o$/s and $250^o$/s, with an average velocity of about $180^o$/s. Figure 13 shows the saccade duration distribution, averaged over all search array sizes, for young and older adults.

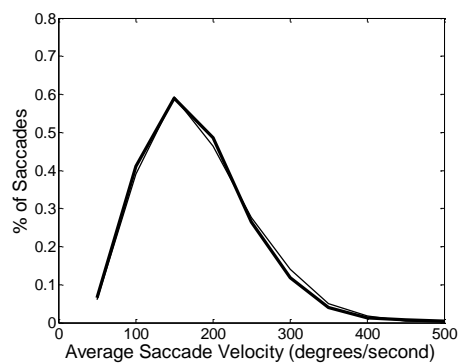

Figure 12. Saccade velocity distributions for young (thick line) and older (thin line) adults. The Mean/Standard Deviation for young (old) are: 178.3/71.6 (181.4/72.4) degrees per second. The velocity of a saccade is obtained as the ratio of the saccade amplitude to its duration, which thus yields the saccade average velocity.

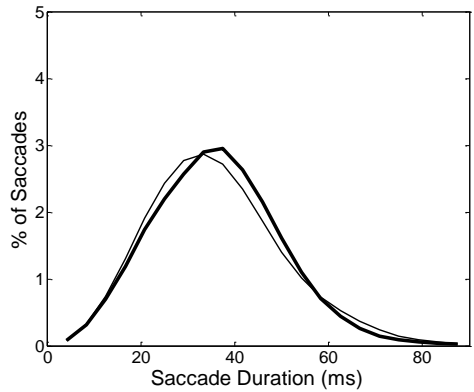

Figure 13. Saccade duration distribution for young (thick line) and older (thin line) adults. The Mean/Standard Deviation for young (old) are: 36.6/14.4 (36.1/14.5) ms.

Overall, there is little difference between young and older adults in these characteristics of the saccades, which are fairly typical for a task like ours, but become much more varied and remarkable for other types of studies and settings —see Aks (2009, p. 139) for some review of nonlinearities in eye movements.

### 3.8. Target Recognition Accuracy

Tables 4(a) and 4(b) show the accuracy of the participants in identifying search arrays with and without the target presence. The two tables differ in that trials in which a participant failed to press any button within the 4-second time limit are excluded in Table 4(a), whereas they are counted as an incorrect response in Table 4(b) —see also failure rates in Sec. 3.1.

The accuracy in target presence/absence identification is comparable for young and older adults. Indeed, there are no significant differences between any pair of corresponding values for young and older subjects reported in either Table 4(a) or Table 4(b). Nevertheless, one may notice that the older participants were able to identify correctly all the trials where the target was absent in 12-distractor arrays. Looking at Table 4(a), it also appears that the

How Do Eyes and Brain Search a Randomly Structured …	27older participants were rather more accurate than the young participants in indicating when the target was absent in 96-distractor arrays. However, older participants may have been merely adopting a more conservative decision criterion, since their latter superiority in accuracy disappears when the timed-out trials are included in Table 4(b).

**Table 4(a). Accuracy of participants in identifying search arrays with and without the target present, excluding timed-out trials.**

| Search Array Size | Young Adults | | Older Adults | |
|---|---|---|---|---|
| | Correctly Indicated Target Present (%) | Correctly Indicated Target Absent (%) | Correctly Indicated Target Present (%) | Correctly Indicated Target Absent (%) |
| 12 | 99.2 | 95.2 | 98.3 | 100 |
| 24 | 97.8 | 93.7 | 97.3 | 93.7 |
| 48 | 94.0 | 93.7 | 90.7 | 96.4 |
| 96 | 84.4 | 85.1 | 84.6 | 95.0 |

### 3.9. Target Detection Distance

Figures 14(a-d) show the target detection distance distributions for young and older adults searching each array size. Namely, the ordinate provides the percentage of trials in which the target was detected at the distance from the last fixation marked in abscissa. It appears that young adults were able to detect the target at greater distances than older adults, particularly for the smaller search array sizes. However, those differences between our age groups again did not reach statistical significance.

**Table 4(b). Accuracy of participants in identifying search arrays with and without the target present, including timed-out trials**

| Search Array Size | Young Adults | | Older Adults | |
|---|---|---|---|---|
| | Correctly Indicated Target Present (%) | Correctly Indicated Target Absent (%) | Correctly Indicated Target Present (%) | Correctly Indicated Target Absent (%) |
| 12 | 99.2 | 95.2 | 98.3 | 100 |
| 24 | 97.8 | 93.7 | 97.3 | 93.7 |

| 48 | 93.1 | 93.7 | 89.5 | 88.7 |
| 96 | 78.4 | 63.3 | 73.9 | 61.3 |

Larger areas of conspicuity for young adults are nonetheless indicated by other measures and supported by the computer simulations of Keech and Resca (2010b). Scialfa et al. (1994) have also inferred a reduction in *useful field of view* in older adults performing a more complex feature search. The *distributions* shown in our Figures 14(a-d) are conceptually related to, but technically different from, the *absolute probabilities* of target capture on the next saccade for any mid-trial fixation ―in which the target could or could not have been recognized― shown in Figure 5(A) of Motter and Belky (1998a, p. 1012). Such decreasing probabilities of target capture with increasing target eccentricity in fact provide a more precise definition of the AC. Figure 5(B) of Motter and Belky (1998a, p. 1012) then shows a remarkable scaling property of such AC, which becomes essentially independent of array size once that distances are expressed in ANND units.

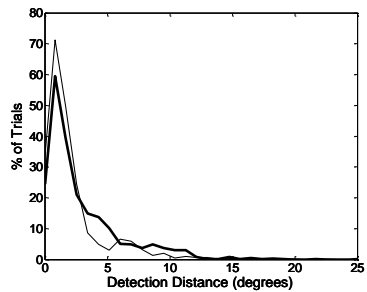

(a) 12-distractors

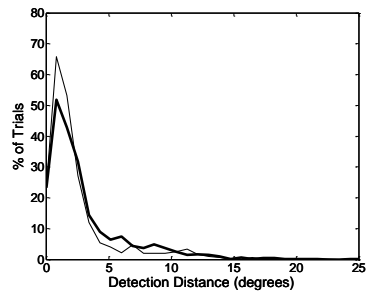

(b) 24-distractors

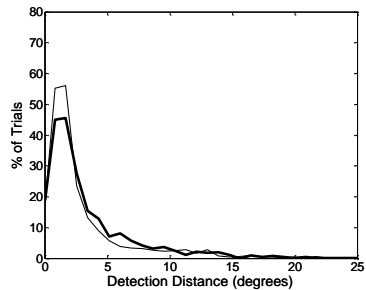

(c) 48-distractors

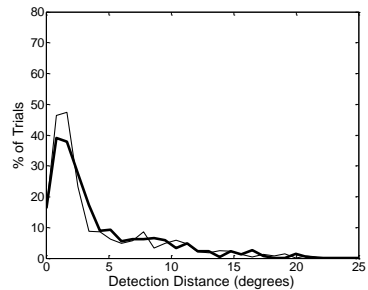

(d) 96-distractors

Figure 14. Target detection distance distributions for young (thick lines) and older (thin lines) adults. For each search array size, the Mean/Standard Deviation for young (old) are: (a) 2.91/2.83 (2.20/2.35); (b) 2.98/2.97 (2.48/2.71); (c) 3.35/3.39 (3.02/3.41); (d) 4.13/4.05 (4.01/4.21).

So, the AC decreases with increasing *stimulus density*, which represents the central point (1) of the Motter-Belky model. In that regard, Figures 5(A) and 5(B) of Motter and Belky (1998a, p. 1012) are clearly more informative than our Figures 14(a-d).

## 4. MEMORY

In this section, we address the central question of whether or what sort of memory is involved in the visual search task that we investigated. We first consider a "survivor function" analysis of *cumulative probabilities* of target capture, which indicates use of memory by our human participants. We then discuss the "design problem" in the study of Motter and Belky (1998a, 1988b), which impedes a corresponding assessment for their monkeys. *Return rates* are then reported, which are also found indicative of some sort of *long-range memory*.

We do not attempt to investigate further in this chapter which memory processes or underlying psycho-neural mechanisms may be more specifically involved in our visual search task. Processes such as inhibitory tagging, systematic scanning or scanpath strategies are instead extensively considered in a study of patterns underlying the eye movement trajectories (Keech and Resca, 2010a) and in a computational model and simulations of the visual search (Keech and Resca, 2010b).

### 4.1. Survivor Function Analysis

In order to investigate whether or not memory for previously fixated locations played a role in a subsequent experiment with monkeys performing a similar conjunctive visual search task, Motter and Holsapple (2001) employed the following "survivor function" approach.

Suppose that the probability of capturing the target is exactly the same for any fixation during the entire search process, namely, a constant $P$ independent of previous fixations. This corresponds to complete absence of memory. If that is the case, the *cumulative probability* $P_C(k)$ of capturing the target after $k$ fixations is precisely

$$P_C(k) = 1 - (1 - P)^k ,  \qquad (1)$$

or, equivalently,

$$Log(1 - P_C(k)) = k \cdot Log(1 - P) = m \cdot k . \qquad (2)$$

Now, Equation 2 is represented by a straight line with a constant negative slope $m = Log(1 - P)$ in a plot of the $k$-dependence of $Log(1 - P_C(k))$. On the other hand, the integral of the relative-frequency distribution of the number of fixations can be determined from observed data and associated with $P_C(k)$ in a corresponding plot of Equation 2. So, if the observed $Log(1 - P_C(k))$ turns out to be basically a linear function of $k$, one may conclude that $P$ is essentially constant, suggesting that memory is practically absent in the search process.

Based on this approach, we show in Figure 15 cumulative probabilities $P_C(k)$ for the young (a) and older (b) participants in our study, searching each type of array sizes. Figures 16(a-b) show the corresponding plots of $Log(1 - P_C(k))$ as a function of the number $k$ of fixations per trial.[6] For comparison, we show in Figures 17(a,b) similar $P_C(k)$ and $Log(1 - P_C(k))$ plots for the two monkeys in the main experiment of Motter and Belky (1998a, 1998b). Namely, in Figure 17(a) we have reproduced as accurately as possible the data for $P_C(k)$ reported in Figure 9 of Motter and Belky (1998b, p. 1812). From those data, we have then computed $Log(1 - P_C(k))$ and reported their values in Figure 17(b), for each search array size.

---

[6] We should also mention that the frequency distributions of the number of saccades, hence, of fixations in Figures 2(a-d) included all target-present trials, even those in which participants failed to detect the target because of the 4-second time limit imposed on the search (see failure rates for various array sizes reported in Sec. 3.1). Probabilities $P_C(k)$ of "capturing" the target after $k$ fixations in Figures 15(a-b) have been obtained by integration of those *relative* frequency distributions. Thus, such "cumulative" probabilities $P_C(k)$ are inherently normalized to approach 1 with increasing $k$, even though in a finite fraction of trials the target could *not* be actually captured. However, we are only interested in determining whether the probability $P$ of target capture at each fixation changes or not as the search progresses, and that is determined solely by the $k$-sequencing of fixations, regardless of whether the target could ultimately be captured within or beyond any arbitrarily restrictive time limit on $k$.

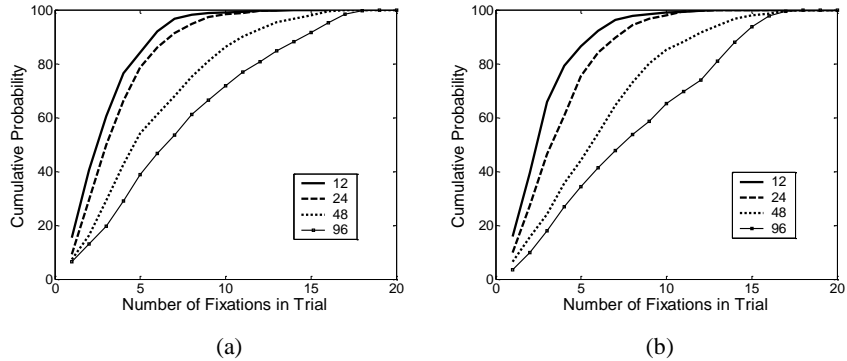

Figure 15. Cumulative probabilities $P_C(k)$ of capturing the target as a function of the number k of fixations in target-present trials for young (a) and older (b) adults.

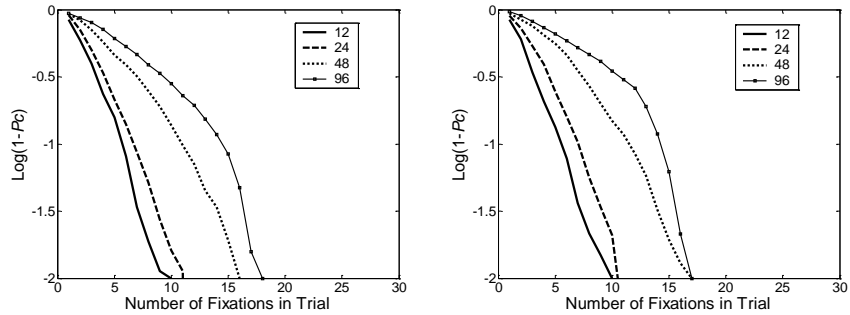

Figure 16. Log(1-$P_C(k)$) as a function of the number k of fixations in a trial for young (a) and older (b) adults.

Our Log-plots in Figures 16(a-b) display *negative curvature* right from the onset, suggesting the *presence of memory*, with a corresponding *increase* in *P* as the search progresses.[7] Vice versa, the Log-plots for monkeys in Figure

---

[7] If *P* is not a constant, but rather a function $P_j$ of the *j*-fixations, we should replace $(1-P)^k$ in Equation 1 with its geometrical average $\left(1-\langle P\rangle_k\right)^k = \prod_{j=1}^{k}\left(1-P_j\right)$. Then the slope $m = Log(1-\langle P\rangle_k)$ of the

17(b) display *positive curvature*, implying a *reduction* in P, as the search progresses. These figures may be further compared with Figure 9.3(B) of Motter and Holsapple (2001, p. 162), which instead displays mostly *linear* Log-plots up to $P_C(k)$ values of about 0.9, corresponding to $Log(1-P_C(k))$ values of about −1. So, which of the three is the "right" curvature behavior?

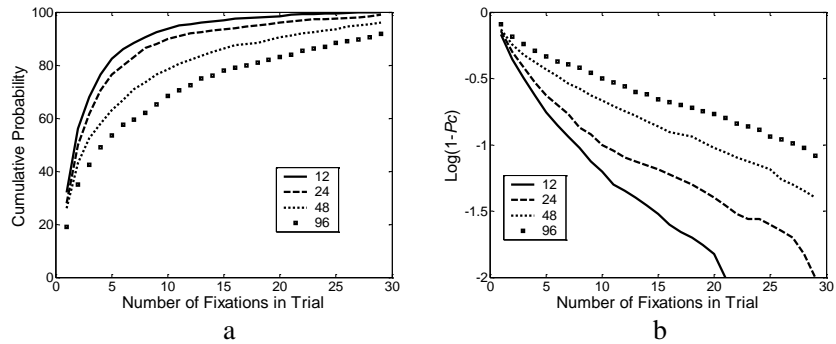

Figure 17. $P_C(k)$ in (a) and $Log(1-P_C(k))$ in (b) are reproduced for monkeys from Fig. 9 of Motter and Belky (1998b).

Taking finite-*k* derivatives of the $P_C(k)$ curves reproduced in Figure 17(a), we derived the distributions of the saccade numbers per trial for the two monkeys in the Motter-Belky main experiment, which are reported in Figure 18.

---

*Log* (1-$P_C(k)$) curve becomes also a function of *k*. If the $P_j$'s are slowly increasing, for example, so is their corresponding average $<P>_k$. In our discussion of curvature trends, we should refer more precisely to the average $<P>_k$ in the case of a varying *P*. See Sec. 4.3 and Equation 3 in particular.

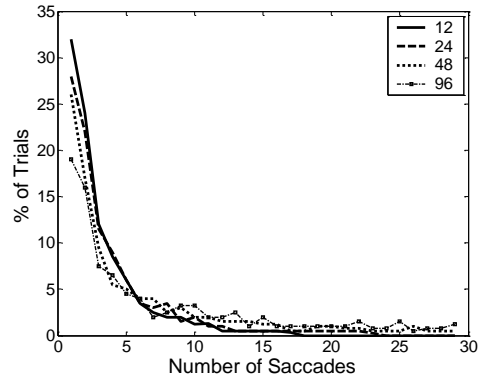

Figure 18. Inferred distributions of the number of saccades per trial for the two monkeys in the Motter-Belky main experiment. The mean/standard deviation for the 12-, 24-, 48-, and 96-distractor arrays are 3.18/2.96, 3.92/4.14, 5.74/6.5, and 6.94/7.44, respectively.

Comparing those with our Figures 2(a,d), it is evident that monkeys and human subjects conducted their respective visual searches quite differently. Unquestionably, the monkeys had a greater chance $P$ of capturing the target quite early in the trial. However, if they did not succeed early, their chance $P$ dropped considerably. So, even though the monkey and human overall performances were rather similar on average, as noted toward the end of Sec. 3.1, the complete saccade number distributions actually differ substantially in monkeys and humans.

## 4.2. Memory or No Memory: A Problem of Design?

While a perfectly constant $P$ implies absence of memory, presence of memory is consistent with an increase in $P$ as the search progresses. But what could cause a *reduction* in $P$ as the search advances, as observed for the monkeys in the main experiment of Motter and Belky (1998a, 1998b)? The answer, most likely, is the design problem in their placement of the target already mentioned in our Introduction. Namely, the procedure that we have described in Sec. 2 to generate the search arrays in our investigation satisfies the basic requirement of a truly random and *uniform* distribution of target locations throughout the display, whereas the procedure described by Motter and Belky (1998a, p. 1008) unfortunately does not.

Incidentally, this design problem is also the most likely explanation of a certain discrepancy in the probabilities of target capture on the next saccade for mid-trial fixations and the initial fixation, shown in Figures 5(B) and 7(B) of Motter and Belky (1998a, pp. 1012-1013). Namely, for the initial fixation at the display center, they found that "first, the probability that a nearby target will be detected and fixated next is greater than for mid-trial fixations, and second the probability gradient for the initial fixation condition is steeper" (p. 1013). Both findings are probably a consequence of the target being more likely placed in central than peripheral regions.

Fortunately, Motter and his collaborators were not pleased with their original array construction, suspecting that it may have created some problems, as we found independently. So, they switched to a different design of their search arrays, in which targets and distractors were drawn truly at random from the same pool of possible stimuli. That change was implemented in the study of Motter and Holsapple (2001), which also involved a somewhat different task of searching through randomly rotated "T and L" distractors and target —thus further eliminating the need for *color filtering*.

### 4.3. Limitations of the Survivor Function Approach and Possible Improvements

If the probability $P_j$ of capturing the target varies at different $j$-fixations along their sequence in a search trial, Equation 1 must be replaced by

$$P_C(k) = 1 - \prod_{j=1}^{k}(1-P_j) \equiv 1 - \left(1 - \langle P \rangle_k\right)^k, \tag{3}$$

where the last step defines a geometric average $<P>_k$ of $P_j$ probabilities up to the $k$-fixation.

Now, with increasing $k$, the $<P>_k$ average may become rather insensitive to fairly rapid $P_j$ variations. Taking logarithms may further reduce sensitivity, making possible, for example, an approximately linear fit of $Log(1-P_C(k))$. That could suggest a fairly constant $<P>_k \approx P$, when in fact some $P_j$ values could differ considerably from that constant $P$. Furthermore, even when it is clear that a linear fit of $Log(1-P_C(k))$ is unwarranted, as in our Figures 16(a-b), there is no immediate statistical procedure to assess the significance of the logarithm curvature and its effects.

Presumably, a more sensitive and informative way of applying the "survivor function" method is to determine directly the probabilities $P_j$ of capturing the target at various $j$-fixations from the observed cumulative probabilities $P_C(k)$ consecutively as

$$P_j = \frac{P_C(j) - P_C(j-1)}{1 - P_C(j-1)} \qquad (4)$$

Motter and Holsapple (2007) have done that recently, showing in their Figure 7(a) $P_j$ probabilities that vary considerably, even though the corresponding $Log(1 - P_C(k))$ curves in their Figures 6(c-d) can still be fitted rather well *linearly* (p. 1273). Carrying out a similar analysis of our $P_j$ values, a comparison could have more sensitively illustrated possible differences between their highly trained monkeys and our relatively untrained humans of different ages.

## 4.4. Rates of Return to Previously Inspected Locations

We plot in Figures 19(a-d) the percentage of saccades that return to any given previous fixation location (within 1.4°) as a function of the intervening number $l$ of saccades, for young and older adults. In all these plots, the largest peak, ranging from about 1% to about 4%, occurs for the shortest delay of just 2 saccades, and is relatively more pronounced for the young adults. This *immediate return rate* ($l$=2) is reasonably consistent with the estimate of Motter and Belky (1998b, p. 1810) in their Figure 7(B) that a return saccade to the location of the preceding fixation occurred for about 3% to 4% of all saccades. As the search progresses beyond that, our Figures 19(a-d) typically show decreasing return rates with increasing delays, meaning fewer and fewer returns to earlier and earlier fixation locations.

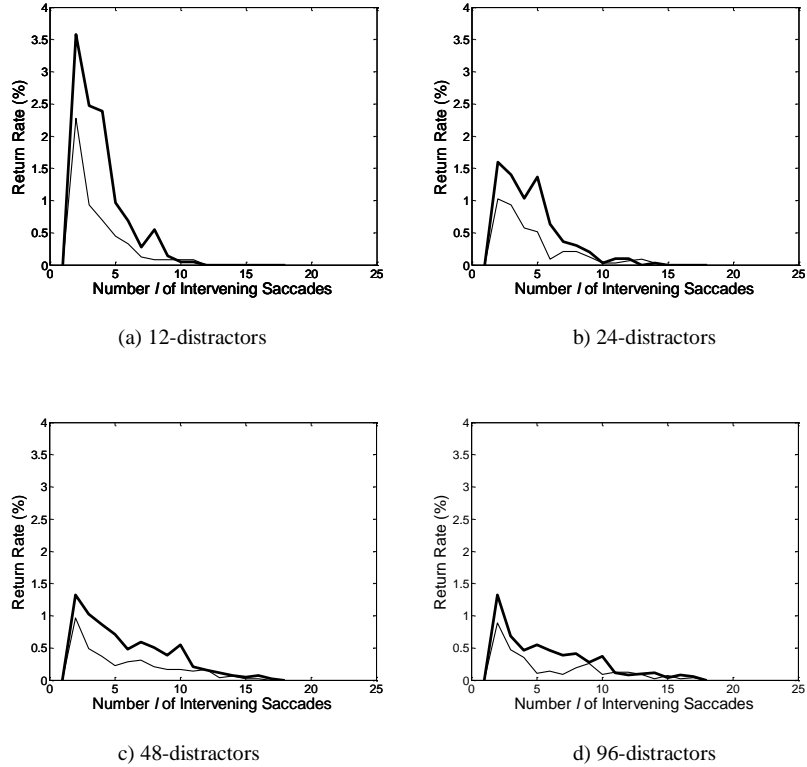

Figure 19. Return rate (percentage) to an original fixation location after l intervening saccades for young (thick lines) and older (thin lines) adults.

To a certain extent, that may be due to the fact that eye trajectories tend to spiral outward from the display center with a forward bias, which tends to progressively reduce the likelihood of return to central locations (Keech and Resca, 2010a, 2010b).

In order to estimate a return *probability*, we must take into account that the number of opportunities to return to an earlier fixation location diminishes as the search progresses. For example, in a trial in which the target was found after 4 fixations, there were 2 opportunities to return to a previous fixation location with a delay of $l=2$ saccades, namely, 3 to 1 and 4 to 2. On the other hand, there was a single opportunity to return to a previous fixation location with a delay of $l=3$ saccades, namely, 4 to 1. Accordingly, we consider the percentage of returns relative to the number of opportunities to return, shown in Figures 20(a-d) for young and older adults.

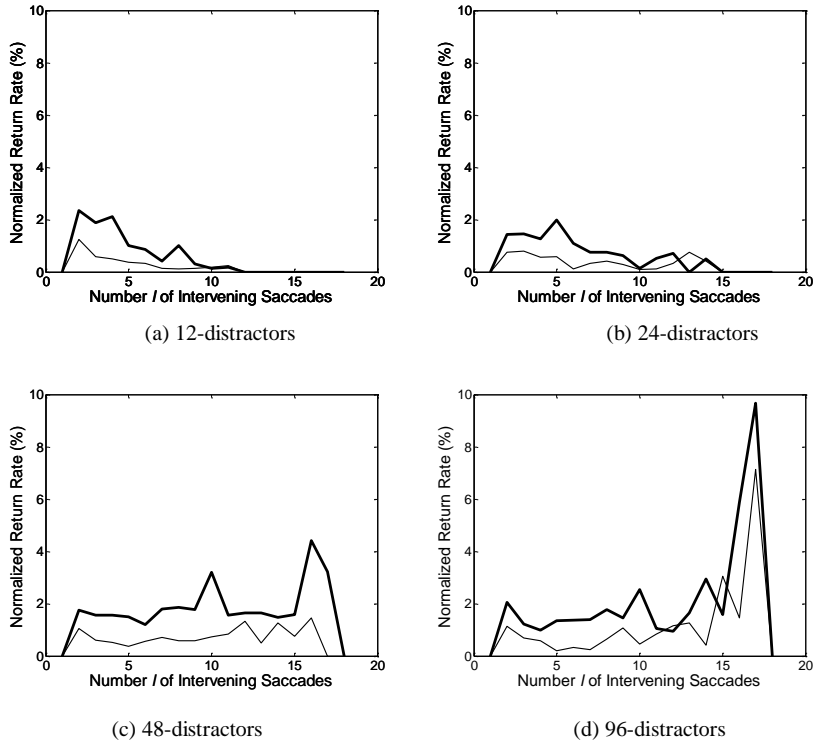

Figure 20. Normalized return rate (percentage) to an original fixation location after *l* intervening saccades for young (thick lines) and older (thin lines) adults.

These *normalized return rates* are only slightly decreasing (for 12- and 24-distractor arrays) or relatively constant (for 48- and 96-distractor arrays) for increasing delays. This suggests that there is substantial *long-range memory* of past fixations in our visual search task.

Only after as many as 15 intervening saccades, which occur rarely and only for 48- or 96-distractor arrays, the normalized return rates finally spike up ―to about 10% and 7% in 96-distractor arrays searched by young and older adults, respectively. Even that is more likely related to unavoidable closure in a few trajectories spiraling outward-and-inward than to any ultimate loss of memory (Keech and Resca, 2010a).

For a *completely random* search, each *relevant* object (distractors and target of the same color) would have the same constant probability 2/N of being selected at each new fixation among all (N+1) objects ―assuming a full color segmentation of the stimuli. That means probabilities of 16.7%, 8.3%, 4.2%, 2.1% for N = 12, 24, 48, 96 distractors in the search arrays. These probabilities are typically much higher (except for 96-distractor arrays) than normalized return rates observed in Figure 20. That may also suggest considerable memory. However, crude probability estimates like 1/N or 2/N are not really applicable to saccadic eye movements, which can hardly be *completely random*. Any realistic model should take into account at least the constraints of the *saccade amplitude distribution*, which is hardly constant for all inter-object distances (Figure 7), and of *the forward bias* (Figure 10(b)), which is hardly negligible. Including those features, computer simulations have indicated that the return rates observed in Figures 19 and 20 cannot be consistently reproduced for *all* delays unless some form of *long-range memory* is also assumed (Keech and Resca, 2010b).

## CONCLUSION

We conducted an experimental study with an eye-tracker on 7 young and 7 older human subjects performing a basic conjunctive visual search task relatively similar to one performed by two rhesus monkeys in an original influential study of Motter and Belky (1998a, 1998b). A detailed analysis of our observations has confirmed all the central elements identified by Motter and Belky (1998a, 1998b) with regard to the roles played by perception, selection, attention, and object recognition in the visual search process. However, our study further indicates a critical role played by long-range memory.

In particular, our observations confirm the following:

(1) During each fixation, human subjects, like monkeys, can process in parallel several items for target recognition, namely those around the fixation point and within a certain radius, which depends on the *local density* of *relevant* stimuli —namely, those sharing *color* with the target, rather than *orientation*. This zone of focal attention has been called *area of conspicuity* (AC) for *target recognition* (Engel, 1971, 1974, 1977; Motter and Belky, 1998a, 1998b), or, alternatively, *functional visual field* (Sanders, 1970), *useful field of view* (Ball et al., 1988; Scialfa et al., 1987; Scialfa et al., 1994), and *visual lobe area* (Kraiss and Knäeuper, 1982) —see Keech and Resca (2010a) for further review of these and other related concepts. Further characterization of the AC and understanding of its neuro-cortical representation have been subsequently provided (Motter and Holsapple, 2000, 2001, 2007; Motter and Simoni, 2007, 2008).

(2) The most probable saccade amplitudes are $5^o$, $5^o$, $5^o$, $4^o$ for young adults searching arrays with 12, 24, 48, 96 distractors, and $4^o$, $4^o$, $3^o$, $3^o$ for older adults, correspondingly (cf. Figure 7). These amplitudes provide a reasonable estimate of the AC radius. Accordingly, young adults have slightly greater conspicuity areas than older adults, especially for larger arrays. The most probable saccade amplitudes decrease with increasing array size. On the other hand, in units of the average distance between *relevant* nearest neighbors (sharing color with the target), or c-ANND, the most probable saccade amplitudes for the young adults increase as 1.0, 1.4, 1.8, 1.9, while the corresponding results for older adults increase as 0.8, 1.1, 1.1, 1.4 (cf. Figure 8). These observations are consistent with corresponding estimates from computer simulations of the eye trajectories performed by Keech and Resca (2010b).

On the other hand, our observations indicate the following:

(3) A "survivor function" analysis of cumulative probabilities of target capture demonstrates that our human participants relied substantially on some form of *long-range memory* to perform our visual search task. Our finding of memory is far from minor, involving, for example, observed return rates smaller by a factor of five or more than those of memoryless simulations over a long range of intervening fixations (Keech and Resca, 2010b), far exceeding current estimates of a rapidly decaying IOR (Snyder and Kingstone, 2000; Hulleman, 2009). Vice versa, deployment of long-range memory was not found by Motter and Belky (1998a, 1998b), a conclusion that Motter and Holsapple (2001) later tested, confirmed and modeled more extensively in a fairly similar visual search task, conducted again with two rhesus monkeys.

In order to understand such discrepancy, we carefully considered every possible design peculiarity or difference. For example, Motter and Belky (1998a, 1998b) generated search arrays with distributions slightly more uniform than ours, particularly for smaller search arrays. However, we found that effect to cause only minor shifts in saccade amplitude distributions ―see Sec. 3.4. Perhaps more significantly, subjects communicated to the experimenter their detection of the target differently in the two sets of experimental designs. In our study, participants were required to make a speeded search decision about the presence or absence of the target in the array by *pressing* one of two buttons on a keyboard. The search array remained on for 4.0 sec or until a response to the search task was made. By contrast, Motter and Belky (1998a, 1998b) trained their monkeys to *fixate* the (always present) target for 600 ms to complete the trial, and their search array stayed on for 7.262 sec or until the target was detected.

With regard to this difference, the "null hypothesis" is that the modality by which the detection of the target is communicated has no influence on how the visual search is conducted by the subject prior to that detection and communication. An "alternative hypothesis" is that the requirement of target *fixation* may affect the deployment of attention from the start, and it may thus alter the way in which the visual search is conducted all along. These hypotheses can be tested by having the visual search task performed twice by each subject, once with either modality of communicating the detection of the target (pressing a button versus target fixation). We investigated that by collecting two sets of data under the two alternative detection conditions for 4 individuals, out of the total of 14 participants in our main study. The analysis of the additional set of data, requiring target fixation, did not show any significant difference in comparison to the original set of data, requiring pressing of a button. That supports the null hypothesis of no influence of the target fixation requirement on the way in which the visual search is conducted all along.

There was a certain design problem in the main experiment of Motter and Belky (1998a, 1998b) with regard to the placement of the target, which was neither random nor uniform, but rather orderly and preferentially central. On the other hand, there was no such problem in a subsequent study of Motter and Holsapple (2001). Yet, their observations of linear Log-plots for the cumulative probabilities of target capture led them to conclude that memory should be relatively limited in the visual search process. We are thus confronted with the possibility that monkeys and humans may *not* use mnemonic resources compatibly already in some basic visual search tasks, perhaps contrary to a common expectation. In fact, the cortical representation of the AC construct does not appear to differ qualitatively in monkeys and humans (Maioli, Benaglio, Siri, Sosta, and Cappa, 2001; Motter and Simoni, 2007; Simoni and Motter, 2003).

On the other hand, there is also the possibility that search strategies, and use of memory in particular, may change over time as a result of extensive practice with a specific visual search task, which was the condition for the highly trained monkeys in the studies of Motter and his collaborators, but not the condition for the relatively untrained human participants in our study. However, this alternative possibility begs the question of how the search could maintain efficiency by forsaking whatever use of long-range memory, which should be most valuable in searching scenes with many completely randomized and uninformative items, like in our and Motter's displays — compare random searching with and without replacement (Arani, Karwan, and Drury, 1984). Perhaps habituation unavoidably leads to a diminished use of precious but exacting mnemonic resources, despite the associated loss of efficiency and reward in performing the task. Or perhaps extensive training induces enhancement or recruitment of some other neural resources to compensate for lacking memory, even though we saw no behavioral evidence nor hint of that in any of our observations, comparisons and analyses.

We have also addressed this central question of memory, its precise long-range form, and some possibly related search strategies in two related articles (Keech and Resca, 2010a, 2010b) —see, also, Dickinson and Zelinsky (2007) and Peterson et al. (2001, 2007). However, a complete resolution of such intriguing questions or understanding of such puzzling behavior may require further comprehensive investigations and comparative analyses of monkeys and human participants performing precisely the same visual search tasks through equivalent progressions in their levels of training.

Finally, we have also attempted to characterize early signs of cognitive aging in eye movements and corresponding psycho-neural mechanisms of the visual search process. However, our sample of 7 young and 7 older adults was relatively too small and the age differential was relatively too limited to produce major differences systematically surpassing standard tests of statistical significance, with the notable exception of a test on the overall *reaction time*. Nevertheless, in their aggregate, our results consistently indicate that even "moderately older" adults (in fact, in a middle age range around 50 years) already show signs of slightly smaller conspicuity areas and thus slightly less processing and memory capacities. Such subtle changes or deficits develop precisely in the direction that one would expect on the basis of the age-related "generalized slowing hypothesis" (Plude and Doussard-Roosevelt, 1989; Scialfa and Joffe, 1997). More extensive studies that will likely be conducted in the future on how aging can impact perceptual, attentional and mnemonic processing in the AC structure ―hence, in its cortical mapping and representation (Motter and Simoni, 2007)— and its conceivable diagnostic implications may thus benefit from consideration and comparisons with the data that we have thoroughly presented and discussed in this chapter.

# **APPENDIX**

**Measurement Errors**

The spatial resolution of the ASL Model 501 eye-tracker corresponds to a visual angle of about 0.5°. We need to take into account this instrumental resolution when we generate spatial-variable distributions, such as the saccade amplitude distributions in Figure 7, the fixation-to-object distance distributions in Figure 9, or the target detection distance distributions in Figure 14. The simple procedure that we adopted is to generate histograms with abscissa intervals of 1-degree width, centered at integer values of degrees. Each saccade is assigned to a corresponding interval, based on its amplitude. Namely, the interval centered at 1 degree comprises saccades with amplitudes between 0.5 and 1.5 degrees, and so on for all the following intervals. Only the beginning interval is different, since its width is only 0.5 degrees, while it is regarded as "centered" at 0 degrees. The value in the ordinate corresponding to the center of each interval in the abscissa gives the percentage of saccades with amplitudes falling within that interval, divided by the interval width (which is 1 degree, except for the 0-degree interval). These ordinates are then connected by straight lines, thus forming a continuous (piecewise linear) distribution. Its subtended area is close to 100, since our "polygonal" histogram typically varies slowly enough to fit closely a standard "rectangular" histogram. Also notice that, since saccades with amplitudes shorter than 1.3 degrees were excluded from further analysis in Sec. 3.4, the ordinate of the interval centered at 1 degree still has practically zero value in saccade amplitude distributions such as those in Figure 7, for example.

The ASL Model 501 eye-tracker samples at rate of 240Hz, corresponding to temporal intervals of 4.1667 ms in which the coordinates of the eye positions are reported. Thus, an instrumental resolution of ± 4.1667 ms must be taken into account when we generate distributions of temporal variables, such as those in Figures 4 and 13. Consider, for example, a fixation duration distribution, as shown in Figure 4. The abscissa axis represents the duration time $\tau$, which is sampled by the instrument at 4.1667 ms intervals. If a change in fixation is detected at a sampling time $t_n = n \times 4.1667$ ms, the preceding fixation could have ended at any time between $t_{n-1}$ and $t_n$. On the other hand, if a new fixation is found to persist from an initial time $t_m$ onward, it could have begun at any time between $t_{m-1}$ and $t_m$. Thus, we have an uncertainty of 8.3333 ms on the *duration* of any measured fixation. The simple procedure that we adopted to account for that uncertainty is to average over contiguous sampling triplets. Namely, if $f_{n-1}, f_n, f_{n+1}$ represent the percentages of fixations reported by the eye-tracker as having contiguous durations $\tau_{n-1} = \tau_n - 4.1667$ ms , $\tau_n$, $\tau_{n+1} = \tau_n + 4.1667$ ms , we consider the average

$$f_n' = [f_{n-1} + f_n + f_{n+1}] / 3, \qquad (A-1)$$

we divide that $f_n'$ by 4.1667 ms, and then we assign that value to the ordinate corresponding to the abscissa at $\tau_n$ —and so on. Finally, we connect adjacent ordinate values with straight lines, thus obtaining a continuous (piecewise linear) fixation duration distribution, as shown in Figure 4, for example. The same procedure is applied to obtain the saccade duration distributions shown in Figure 13. There are various other more or less equivalent procedures of smoothing or filtering sampling data with a time resolution of 8.3333 ms, of course. We tried several, and we obtained qualitatively similar distributions, as expected.

Finally, for saccade velocity distributions, such as those in Figure 12, the uncertainty in the ratio of the spatial to the temporal variable is estimated to be $25°/s$ , having assumed the first-order approximation for the variance of the ratio of the two random variables (cf. Bulmer, 1979, Problem 5.7, p. 79), with an average saccade amplitude of $6.5°$ and an average saccade duration of 36 ms. Our procedure to generate histograms is then applied with intervals of $50°/s$ , centered at integer multiples of $50°/s$. Only the beginning interval is different, its width being only $25°/s$ . Having used $40°/s$ as a velocity threshold below which a "fixation" was declared and above which a "saccade" was declared, there is no contribution within the first $25°/s$ interval. The value in the ordinate corresponding to $50°/s$ in the abscissa gives the percent of saccades with average velocities falling between $25°/s$ (nominally, but $40°/s$ actually) and $75°/s$, divided by the interval width of $50°/s$. We then proceed similarly for the subsequent $50°/s$ intervals in abscissa. The interval-centered ordinates are then connected by straight lines, thus forming a continuous (piecewise linear) distribution.

How Do Eyes and Brain Search a Randomly Structured … 45